\documentclass[conference]{IEEEtran}
\IEEEoverridecommandlockouts
\usepackage{tikz}
\usepackage{cite}
\usepackage{amsmath,amssymb,amsfonts}
\usepackage{algorithmic}
\usepackage{graphicx}
\usepackage{textcomp}
\usepackage{tabularx}
\usepackage{xcolor}
\def\BibTeX{{\rm B\kern-.05em{\sc i\kern-.025em b}\kern-.08em
    T\kern-.1667em\lower.7ex\hbox{E}\kern-.125emX}}

\begin{document}

\title{In-vehicle Sensing and Data Analysis for Older Drivers with Mild Cognitive Impairment\\

{\footnotesize \textsuperscript{*}Corresponding Author}
\thanks{This research was supported by the U.S. National Institute of Health (NIH) (R01 AG068472 02S1) and National Science Foundation (NSF) (OAC 1948066).}
}

\author{\IEEEauthorblockN{1\textsuperscript{st} Sonia Moshfeghi}
\IEEEauthorblockA{\textit{College of Engg and Computer Science } \\
\textit{Florida Atlantic University}\\
Boca Raton, USA\\
smoshfeghi2021@fau.edu}
\and
\IEEEauthorblockN{2\textsuperscript{nd} Muhammad Tanveer Jan}
\IEEEauthorblockA{\textit{College of Engg and Computer Science} \\
\textit{Florida Atlantic University}\\
Boca Raton, USA\\
mjan2021@fau.edu}
\and
\IEEEauthorblockN{3\textsuperscript{rd} Joshua Conniff}
\IEEEauthorblockA{\textit{Charles E. Schmidt College of Science} \\
\textit{Florida Atlantic University}\\
Boca Raton, USA\\
jconniff@fau.edu}
\and
\IEEEauthorblockN{4\textsuperscript{th} Seyedeh Gol Ara Ghoreishi}
\IEEEauthorblockA{\textit{College of Engg and Computer Science} \\
\textit{Florida Atlantic University}\\
Boca Raton, USA\\
sghoreishi2021@fau.edu}
\and
\IEEEauthorblockN{5\textsuperscript{th}* Jinwoo Jang}
\IEEEauthorblockA{\textit{College of Engg and Computer Science} \\
\textit{Florida Atlantic University}\\
Boca Raton, USA\\
jangj@fau.edu}
\and
\IEEEauthorblockN{6\textsuperscript{th} Borko Furht}
\IEEEauthorblockA{\textit{College of Engg and Computer Science} \\
\textit{Florida Atlantic University}\\
Boca Raton, USA\\
bfurht@fau.edu}
\and
\IEEEauthorblockN{7\textsuperscript{th} Kwangsoo Yang}
\IEEEauthorblockA{\textit{College of Engg and Computer Science} \\
\textit{Florida Atlantic University}\\
Boca Raton, USA\\
yangk@fau.edu}
\and
\IEEEauthorblockN{8\textsuperscript{th} Monica Rosselli}
\IEEEauthorblockA{\textit{Charles E. Schmidt College of Science} \\
\textit{Florida Atlantic University}\\
Boca Raton, USA\\
mrossell@fau.edu}
\and
\IEEEauthorblockN{9\textsuperscript{th} David Newman}
\IEEEauthorblockA{\textit{Christine E. Lynn College of Nursing} \\
\textit{Florida Atlantic University}\\
Boca Raton, USA\\
dnewma14@health.fau.edu}
\and
\IEEEauthorblockN{10\textsuperscript{th} Ruth Tappen}
\IEEEauthorblockA{\textit{Christine E. Lynn College of Nursing} \\
\textit{Florida Atlantic University}\\
Boca Raton, USA\\
rtappen@health.fau.edu}
\and
\IEEEauthorblockN{11\textsuperscript{st} Dana Smith}
\IEEEauthorblockA{\textit{College of Engg and Computer Science } \\
\textit{Florida Atlantic University}\\
Boca Raton, USA\\
danasmith2021@fau.edu}
}

\maketitle

\IEEEpubidadjcol

\IEEEpubid{%
  \begin{tikzpicture}[remember picture,overlay]
    % Calculate the center of the text width
    % Adjust the 'yshift' as necessary to move the box up or down
    \node[anchor=south, align=center, yshift=1cm] at (current page.south) {
      \fbox{
        \parbox{2\columnwidth}{
          \scriptsize
          \copyright 2023 IEEE. Personal use of this material is permitted. Permission from IEEE must be obtained for all other uses, in any current or future media, including preprinting/republishing this material for advertising or promotional purposes, creating new collective works, for resale or redistribution to servers or lists, or reuse of any copyrighted component of this work in other works.
        }
      }
    };
  \end{tikzpicture}%
}

\begin{abstract}
Driving is a complex daily activity indicating age and disease-related cognitive declines. Therefore, deficits in driving performance compared with ones without mild cognitive impairment (MCI) can reflect changes in cognitive functioning. There is increasing evidence that unobtrusive monitoring of older adults’ driving performance in a daily-life setting may allow us to detect subtle early changes in cognition. The objectives of this paper include designing low-cost in-vehicle sensing hardware capable of obtaining high-precision positioning and telematics data, identifying important indicators for early changes in cognition, and detecting early-warning signs of cognitive impairment in a truly normal, day-to-day driving condition with machine learning approaches. Our statistical analysis comparing drivers with MCI to those without reveals that those with MCI exhibit smoother and safer driving patterns. This suggests that drivers with MCI are cognizant of their condition and tend to avoid erratic driving behaviors. Furthermore, our Random Forest models identified the number of night trips, number of trips, and education as the most influential factors in our data evaluation.
\end{abstract}

\begin{IEEEkeywords}
in-vehicle sensing, telematics data, machine learning, cognitive impairment
\end{IEEEkeywords}

\section{Introduction}
{T}{he} World Health Organisation (WHO) and National Institute of Health (NIH) reports reveal that the overall number of people aged 65 and above quickly expands due to improved health care and higher life expectancy. The predictions show that by 2050, the percentage of the world's population aged 65 and up will have increased from 10 percent in 2022 to 16 percent. About 19 percent of the people in North America were aged 65 or older in 2022, making it one of the regions with the highest proportion of older residents~\cite{he2016aging,united2021world}. Mild cognitive impairment (MCI) is an anomalous decline in cognitive function that exceeds the predicted average decline. The clinical concerns persist in differentiating MCI from otherwise healthy conditions in older people~\cite{seifallahi2022detection}. A person with MCI has a decline in cognitive abilities like memory, language, and reasoning skills, having difficulty remembering recent events or conversations, finding the right words, planning, and organizing tasks. Although it is a condition that occurs between healthy brain aging and dementia, they can still perform their usual daily activities. Some cases of MCI may progress into dementia over time. MCI can be caused by various factors, including age-related changes and certain medical conditions, such as Alzheimer's disease.

According to the Veterans Administration's work group on Driving Safety for Veterans with Dementia, older adults with moderate to severe dementia should limit their driving due to safety concerns~\cite{us2016vha}. Several studies describe how emotions in daily life can affect human behavior and facial expression ~\cite{altaher2020using}. Driving is a complex diurnal life activity and can indicate age- and disease-related cognitive declines~\cite{brown2004driving,gold2012examination,rajan2013disability}. Therefore, driving performance deficits compared with those without mild cognitive impairment (MCI) can reflect changes in cognitive functioning~\cite{teasdale2016older}. Research indicates that older people with mild to severe dementia have limited physical and mental abilities~\cite{seifallahi2022detection}. Determining age- and cognitive-related `fitness-to-drive' remains unspecified and controversial since driving licensee varies by state. Current driver evaluation programs can test only a small number of drivers with cognitive concerns, missing many who need to know if they can continue to drive safely. Importantly, no concrete scientific approaches have been investigated to directly understand the current or former status of cognitively impaired older drivers in daily life~\cite{croston2009driving}.

Advances in wireless technology, local computing, and data science can contribute to unobtrusively monitoring daily activities to assess real-world cognitive and functional changes~\cite{kaye2011intelligent,lyons2015pervasive,teipel2018use}. Sensing methods and monitoring tools were developed and applied to real-time data to assess patients' well-being~\cite{shuqair2022incremental}. The potential of using in-vehicle sensing to identify cognitive impairment has recently been studied~\cite{seelye2017passive}. In-vehicle sensing is a novel and unobtrusive idea for measuring driver behavior and detecting cognitive deterioration.
In-vehicle sensors and their applications for monitoring driver behavior have advanced significantly. Some vehicular and behavioral strategies provide viable and inexpensive solutions for diagnosing early dementia in older drivers based on driving characteristics.~\cite{jan2023methods,jan2022non}. Older drivers show age-related changes in driving behavior, such as traveling at a lower speed and self-regulating driving~\cite{adler2005older}. The effects of aging on driving safety prompted the development of a comprehensive program to examine individual and medical information and to conduct behavioral analysis for operational and GPS-based data~\cite{li2017longitudinal}. Machine Learning methods have roles in validating the interaction between the brain function and sensors ~\cite{khademi2022comprehensive,neghabi2022novel}. As a supervised machine learning method, Random Forests demonstrated higher accuracy in identifying and reporting anomalies ~\cite{alkanjr2023iobt}. Real-life driving data with machine-learning approaches for demographic and behavioral factors show that age, number of trips, race/ethnicity, time, and the number of hard brakes have been the most accurate predictors of MCI and dementia in older drivers~\cite{lyu2022using}. Some machine learning models could predict the most critical diagnostic factors, such as Alzheimer’s disease genetics and individual and behavioral patterns for older drivers~\cite{bayat2021gps}.
However, high-quality sensing data and novel scientific techniques that can distinguish cognitive-decline-related changes from age-related ones to understand the impact of cognitive impairment on driving behavior is missing~\cite{eby2012driving}. Therefore, the authors have developed new programmable in-vehicle sensing units with unobtrusive vision and telematics sensors to track and record the subtle driver behavior of older drivers. The novelty of the telematics units used in this study is that they build on microcomputers and enable programmable and expendable hardware and software capabilities. The in-vehicle sensors installed 150+ older drivers' vehicles since 2022 and tracked changes in older drivers' cognitive functioning in a daily-life driving setting. Our programmable in-vehicle sensing units comprise a set of unobtrusive vision and telematics sensors. The recorded changes in driver behavior are compared to results from a battery of cognitive tests (executive function, visuospatial, memory, and language) selected for sensitivity to early subtle changes in cognition and ability to predict driver risk. The collected data for the group of participants under the study was analyzed and visualized using statistical ~\cite{konjalwar2023demonstrating} and machine learning methods.

\begin{figure}[htbp]
\centerline{\includegraphics[width=1\columnwidth]{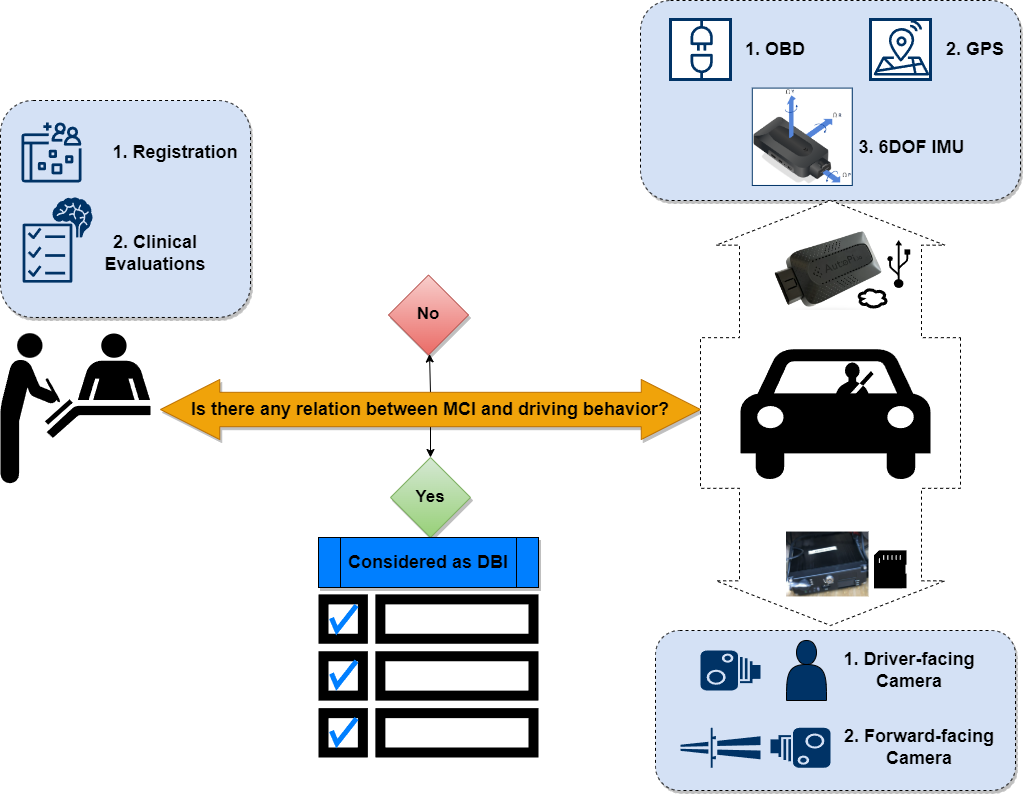}}
\caption{In-vehicle Sensing System}
\label{fig1}
\end{figure}

\begin{figure}[htbp]
\centerline{\includegraphics[width=0.7\columnwidth]{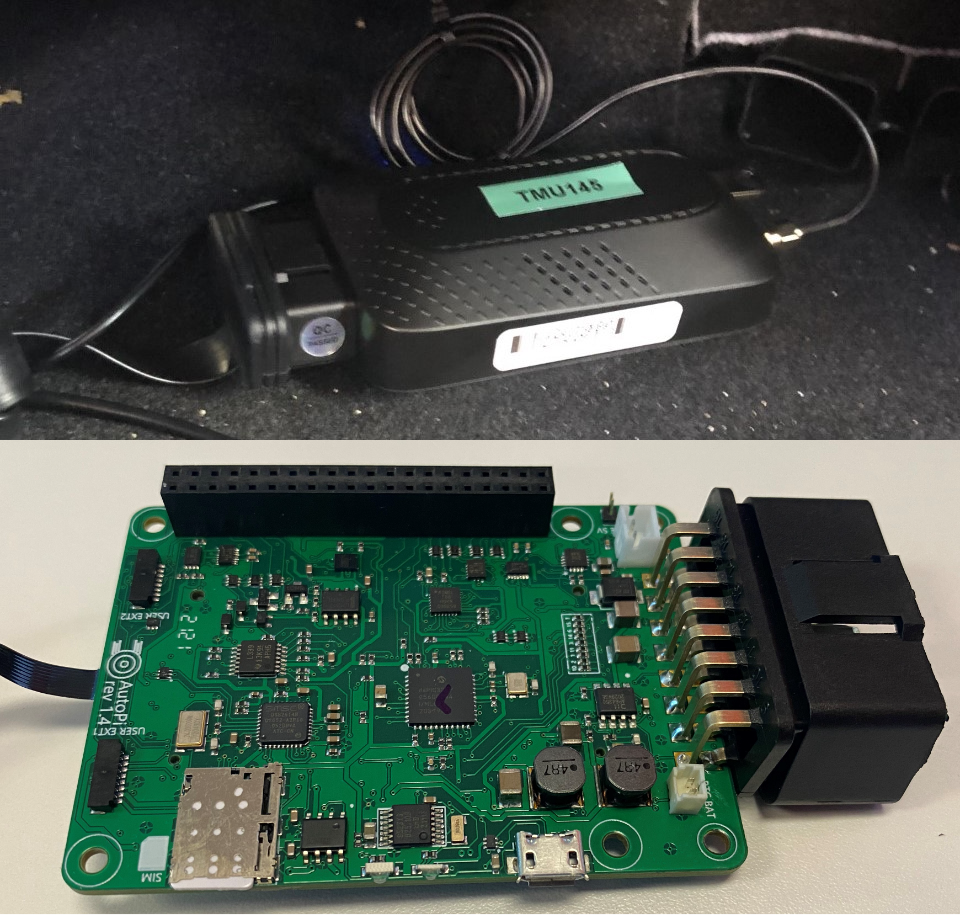}}
\caption{Telematics Unit (TMU)}
\label{fig2}
\end{figure}
\vspace{-10pt}

\section{Design of In-vehicle Sensing}
The authors developed a novel in-vehicle sensing system, which includes telematics and vision sensing units, to collect driver-behavior-related datasets unobtrusively (see Fig.~\ref{fig1}). The in-vehicle sensing system is installed on participants' vehicles for three years to collect longitudinal data streams to capture changes in driving behavior over time. The five-year study was conducted at Florida Atlantic University and the University of Central Florida (UCF) for older drivers with or without MCI and at least 65 with a valid driver's license. The study involved 236 participants in the third year to evaluate drivers' cognitive and behavioral status. All participants completed the initial evaluation form and provided written approval. Individual sessions are subsequently organized every three months to examine any changes in drivers' health and vehicle conditions and to complete related clinical tests, including the Montreal Cognitive Assessment (MoCA), Loewenstein-Acevedo Scales of Semantic Interference and Learning (LASSI-L).

\begin{table}[]
\caption{DBIs Definition and Statistical Overview}
\label{table1}
\resizebox{\columnwidth}{!}{%
\begin{tabular}{cccccc}
\hline
\multicolumn{6}{c}{\textbf{| Driver Indexes |}} \\
\hline
Index & Description & \multicolumn{4}{c}{Statistical Remarks} \\
\cline{3-6}
 & & Min & Max & Mean & StDev \\
\hline
MCI & \begin{tabular}[c]{@{}c@{}}The diagnosis for MCI\\ (1: MCI; 0: non-MCI)\end{tabular} & 0 & 1 & - & - \\
\hline
Age & The participants with 65 years old and above & 65 & 89 & 75.67 & 6.04 \\
\hline
Gender & Male; Female & - & - & - & - \\
\hline
Race & \begin{tabular}[c]{@{}c@{}}1:Black; 2:White; 3:Asian;\\ 4:Native American;\\ 5:Multiracial; 6:Others\end{tabular} & - & - & - & - \\
\hline
Ethnicity & \begin{tabular}[c]{@{}c@{}}1:African-American;\\ 2:European-American;\\ 3:Hispanic-American;\\ 4:Afro-Caribbean;\\ 5:Asian-American;\\ 6:Others\end{tabular} & - & - & - & - \\
\hline
Education & \begin{tabular}[c]{@{}c@{}}Grade school; High school;\\ Some college; Technical school;\\ Associate degree; Bachelor’s degree;\\ Post-graduate studies; Master’s degree;\\ Post-master’s degree; Doctoral degree\end{tabular} & - & - & - & - \\
\hline
Retired & No; Yes & - & - & - & - \\
\hline
BMI & \begin{tabular}[c]{@{}c@{}}Body Mass Index\\ (Obese: greater than 30;\\ non-Obese: lesser than 30)\end{tabular} & - & - & - & - \\
\hline
\multicolumn{6}{c}{\textbf{| Driving Indexes |}} \\
\hline
Total Trip & \begin{tabular}[c]{@{}c@{}}Unit: No.\\ The total number of trips\end{tabular} & 5 & 693 & 266.57 & 174.81 \\
\hline
Night Trip & \begin{tabular}[c]{@{}c@{}}Unit: No.\\ The number of trips at night\end{tabular} & 0 & 58 & 5.8 & 11.56 \\
\hline
Peak Trip & \begin{tabular}[c]{@{}c@{}}Unit: No.\\ The number of trips at peak hours\\ (7:00 to 9:00 and 16:00 to 18:00)\end{tabular} & 0 & 596 & 222.97 & 156.09 \\
\hline
Duration & \begin{tabular}[c]{@{}c@{}}Unit: s\\ The duration of each trip per second\end{tabular} & 0 & 38658.26 & 1280.95 & 2029.02 \\
\hline
Distance & \begin{tabular}[c]{@{}c@{}}Unit: km\\ The distance of each trip per kilometer\end{tabular} & 0 & 412.35 & 10.59942 & 20.55902 \\
\hline
Speed & \begin{tabular}[c]{@{}c@{}}Unit: km/h\\ The speed in kilometers per hour\end{tabular} & 0 & 158 & 26.74934 & 10.99516 \\
\hline
RPM & The revolution per minute & 0 & 5698.25 & 1118.675 & 194.8706 \\
\hline
nH-Acceleration & \begin{tabular}[c]{@{}c@{}}Unit: No.\\ Number of harsh-acceleration\\ (The number of events with an acceleration\\ greater than 3.943 \(m/s^2\) on the X axis)\end{tabular} & 1 & 233 & 3.005261 & 6.084068 \\
\hline
nH-Braking & \begin{tabular}[c]{@{}c@{}}Unit: No.\\ Number of hard-Braking\\ (The number of events with an acceleration\\ greater than -3.943 \(m/s^2\) on the X axis)\end{tabular} & 1 & 123 & 3.614141 & 9.438562 \\
\hline
nH-Turn & \begin{tabular}[c]{@{}c@{}}Unit: No.\\ Number of hard-turns\\ (The number of events with an acceleration\\ greater than -3.943 \(m/s^2\) on the Y axis)\end{tabular} & 1 & 51 & 2.506608 & 2.671595 \\
\hline
Urban Trip & \begin{tabular}[c]{@{}c@{}}Unit: No.\\ The number of trips below 32 kilometers\end{tabular} & - & 7742 & - & - \\
\hline
Suburb Trip & \begin{tabular}[c]{@{}c@{}}Unit: No.\\ The number of trips above 32 kilometers\end{tabular} & - & 38 & - & - \\
\hline
\end{tabular}%
}
\end{table}

\subsection{Programmable Telematics Units}

This study leverages TMUs developed by AutoPi, which build on Raspberry Pi 4 Model B. Fig.~\ref{fig2} shows an overview of the Raspberry-based TMU units. These programmable and open-sourced TMUs support the robust system expandability of hardware and software, allowing the development of customized TMUs to adjust and modify the sensing parameters (e.g., data collection algorithms and sampling frequencies). Each TMU comprises 1) a GPS sensor, 2) an inertial measurement unit (IMU), 3) an OBD connector, 4) a 4G/LTE cellular modem, 5) an SD card, and 6) a USB flash drive. Tri-axial gyroscopes are added to the TMUs to capture angular velocity. It continuously monitors the voltage levels from the port. When the voltage goes down to 12V, it automatically goes into sleep mode for five minutes. After waiting for another 5 minutes, the device goes into hibernate mode. The power usage of the AutoPi in its sleep and hibernation mode is as follows: around 30mA and 10mA.

In-vehicle data can be collected from the Controller Area Network (CAN bus)~\cite{farsi1999overview,johansson2005vehicle} and augmented sensor hardware. Within a group of vehicles, CAN allows the engine control unit (ECU), transmission control unit, brake system control unit, steering system control unit, and others to exchange information quickly and efficiently, providing access to vehicle status data such as sensor readings, control signals, and diagnostic information. One of the most common ways of obtaining CAN bus data is using On-Board Diagnostics (OBD) connectors. Available OBD data includes engine RPM, vehicle speed, and fuel system status. In addition to the OBD data, additional data can be collected from augmented sensor hardware, such as GPS and inertial measurement unit (IMU) modules. Furthermore, various unobtrusive sensing units can be installed on a vehicle to understand the state of drivers~\cite{leonhardt2018unobtrusive}.

\section{Driver Behavior Indexes}
This study used 7794 data points with 19 indexes over two years (see Table~\ref{table1}). The indexes included 19 independent variables as inputs denoted by the matrix of \(X\in\mathbb{R}^{m\times n}\), given by Eq.~\ref{1} to predict the presence or absence of MCI as the dependent or output variable. Considering the values 0 for the absence of MCI (non-MCI) and 1 for the presence of MCI as \(y_i\in\{0,1\}\) for \(i=1,...,m\), the vector of \(Y\in\mathbb{R}^m\) has been obtained. $m$ represents the number of samples, and $n$ is the number of independent variables. Therefore, the values of $m$ and $n$ are 7794 and 19, respectively.

\begin{equation}\label{1}
 X = 
    \begin{bmatrix}
    x_{11} & \cdots & x_{1n} \\
    \vdots & \ddots & \vdots \\
    x_{m1} & \cdots & x_{mn}
    \end{bmatrix}{and}~~~Y = 
    \begin{bmatrix}
    y_{1} \\
    \vdots \\
    y_{m}
    \end{bmatrix}
\end{equation}

Some indexes refer to drivers' characteristics and health circumstances, such as age, gender, race, ethnicity, education, and body mass index (BMI). Both males and females were required to be at least 65 years old. We examined their race/ethnicity to see how their physical traits, nationality, religion, linguistics, or culture influence their driving patterns. To define race/ethnicity, we evaluated six categories for each. The education and employment status of the participants can reveal information about their social, psychological, and economic situation. We divided them into ten groups based on their education level and whether they are employed or retired. Body mass index (BMI) is a unit for calculating body fat by the height and weight of individuals. We classified participants into obese and non-obese groups. We defined a set of indicators for driving features. We studied the total number of trips, the number of trips between 21:00 and 5:00 as night trips, and the number of trips from 7:00 to 9:00 and 16:00 to 18:00 as peak hours trips. Driving characteristics are kinematic factors such as trip duration and distance, speed, rpm, and acceleration. The number of harsh acceleration, hard braking, and hard turns was determined. When the acceleration on the X-axis reaches 3.943 m/s$^2$, it is called harsh acceleration; when the acceleration on the X-axis surpasses -3.943 m/s$^2$, it is deemed hard braking. A hard turn is defined as an acceleration greater than -3.943 m/s$^2$. We evaluated the types of trips following their mileage. Urban trips are defined as distances between 0 and 32 km, whereas suburban trips have a distance greater than 32 km ~\cite{shrestha2017factors}.

\section{Data Analysis}
The in-vehicle data are further analyzed to characterize driver behavior indexes (DBIs) that can reflect physical and cognitive functions over time. 

\subsection{Data Prepossessing}
We combined the quantile normalization method to normalize data distribution and quantile-based flooring and capping to treat outliers, with flooring for lesser values (10th percentile) and capping for higher values (90th percentile). The skewness value dropped for all features after applying this method (e.g., nH-Acceleration decreased from 15.33 to 1.56.).

\subsection{Machine Learning Technique}
We used a supervised machine learning algorithm named Random Forests (RF) for classification purposes. This ensemble learning method integrates multiple machine learning models to improve their performance. Numerous Random Forest algorithm decision trees have been built on a different random subset of training data and features. Each training tree is to predict the target variable, and the final prediction is based on the mode of all the trees' predictions. The random forest training algorithm extends the generic bagging technique to tree learners. Bagging repeatedly ($b$ times) has a random selection with the substitute of the training set and fits trees to these samples, given a training set \(X = x_1, ..., x_i\) with responses \(Y = y_1, ..., y_i\) for \(B = 1, ..., b\), where $i$ is the number of training samples and $b$ shows the number of trees. We developed Random forest classification models with multiple decision trees of (1) only age; (2) the number of trips (for total, peak hours, and night hours); (3) driver variables; (4) driving variables; (5) age with driving variables (6) All the variables to predict MCI status. These six groups show the contribution of each variable in predicting MCI status and the model's performance.
\vspace{-17pt}
\begin{figure}[htbp]
\centerline{\includegraphics[width=0.8\columnwidth]{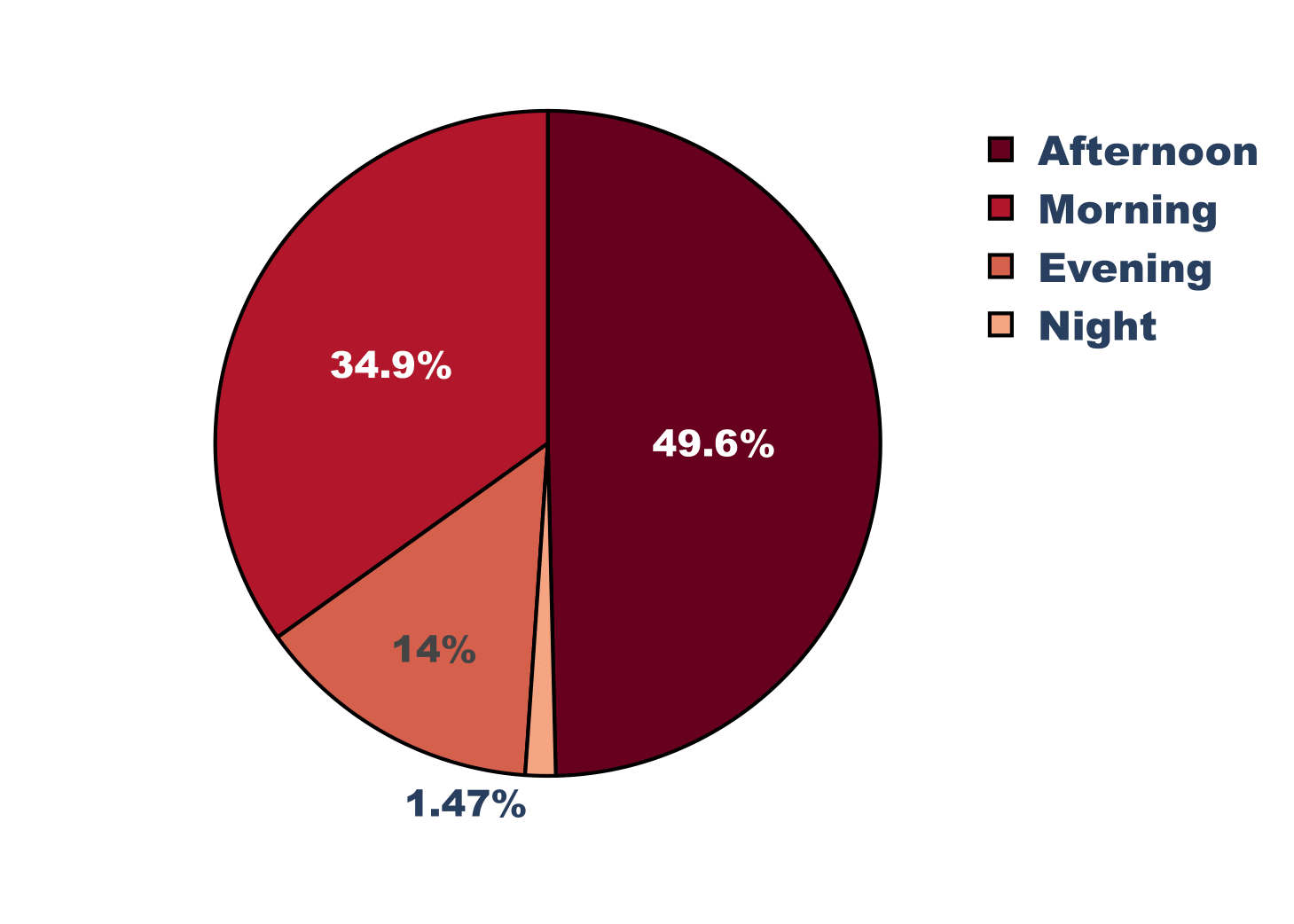}}
\vspace{-12pt}
\caption{Trip Distribution based on Time of Day}

\label{fig3}
\end{figure}

\begin{figure*}[htbp]
    \centering
    \includegraphics[width=0.17\textwidth]{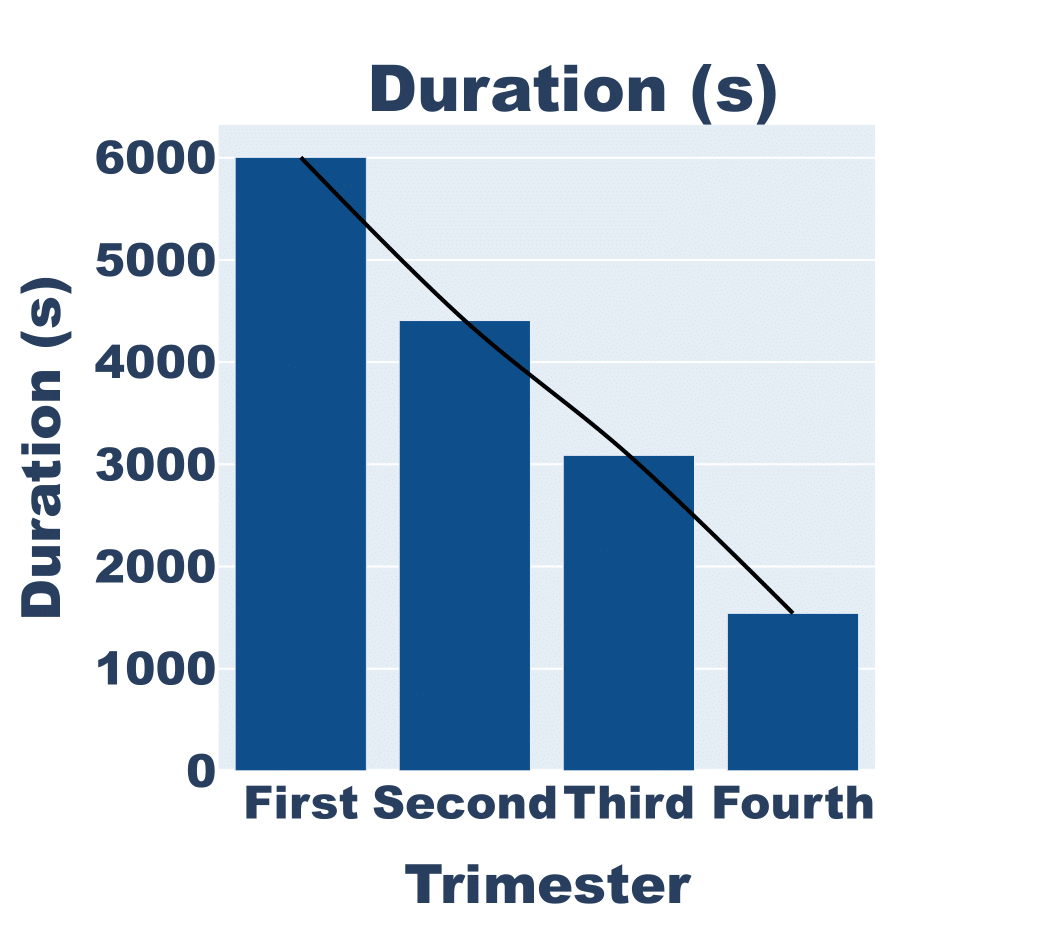}
    \includegraphics[width=0.17\textwidth]{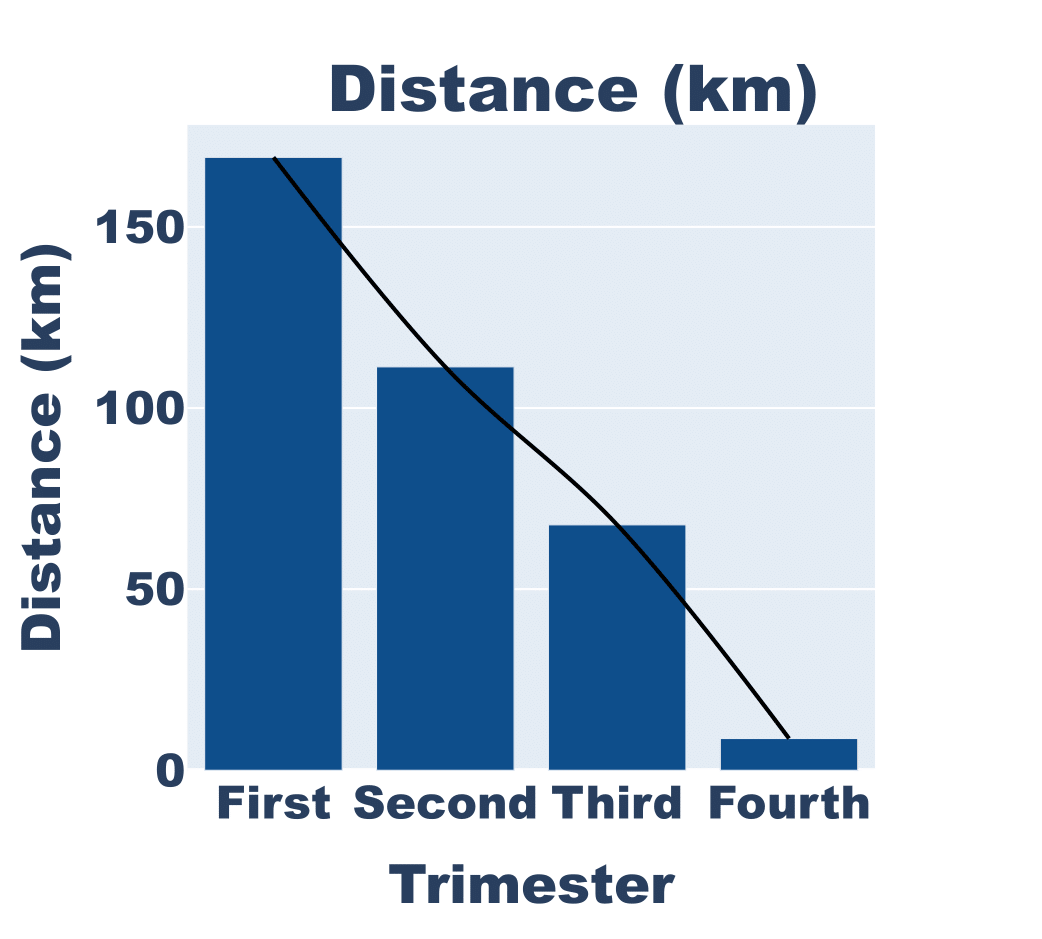}~
    \vline
    \vspace{-10pt}
    \includegraphics[width=0.17\textwidth]{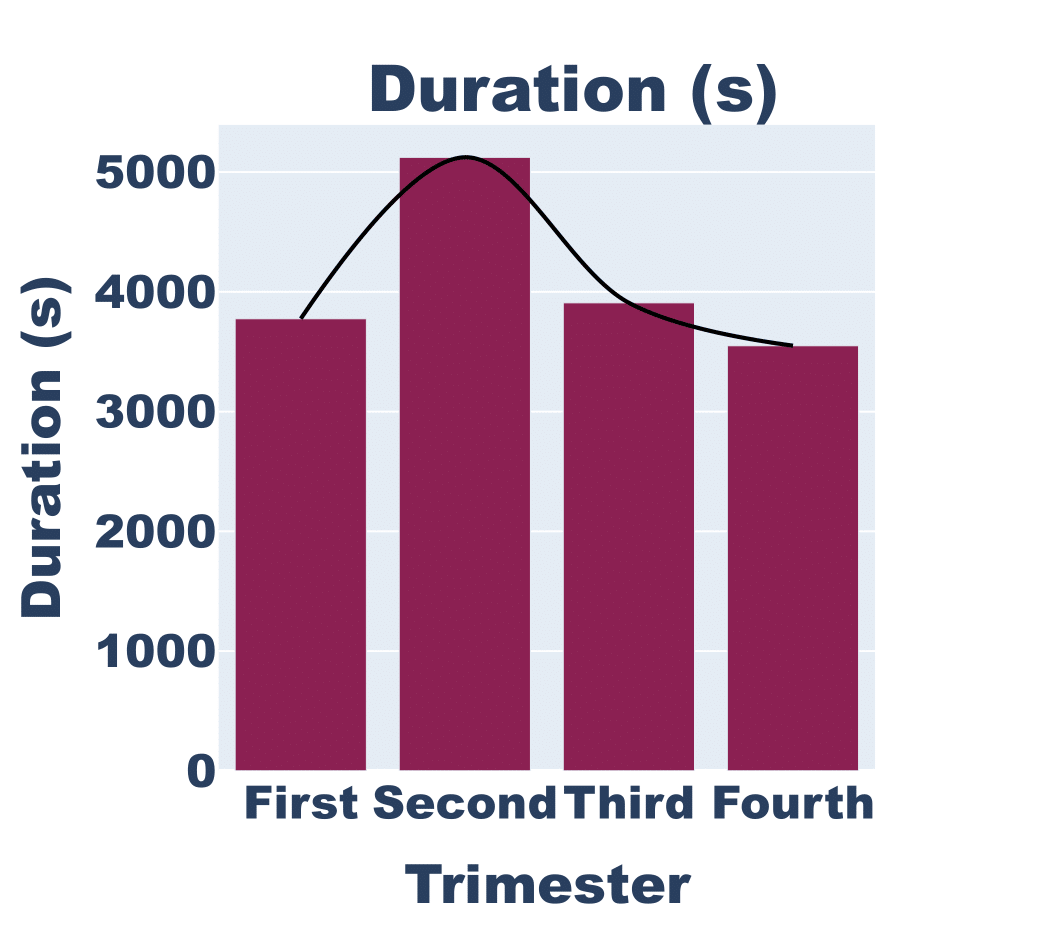}
    \includegraphics[width=0.17\textwidth]{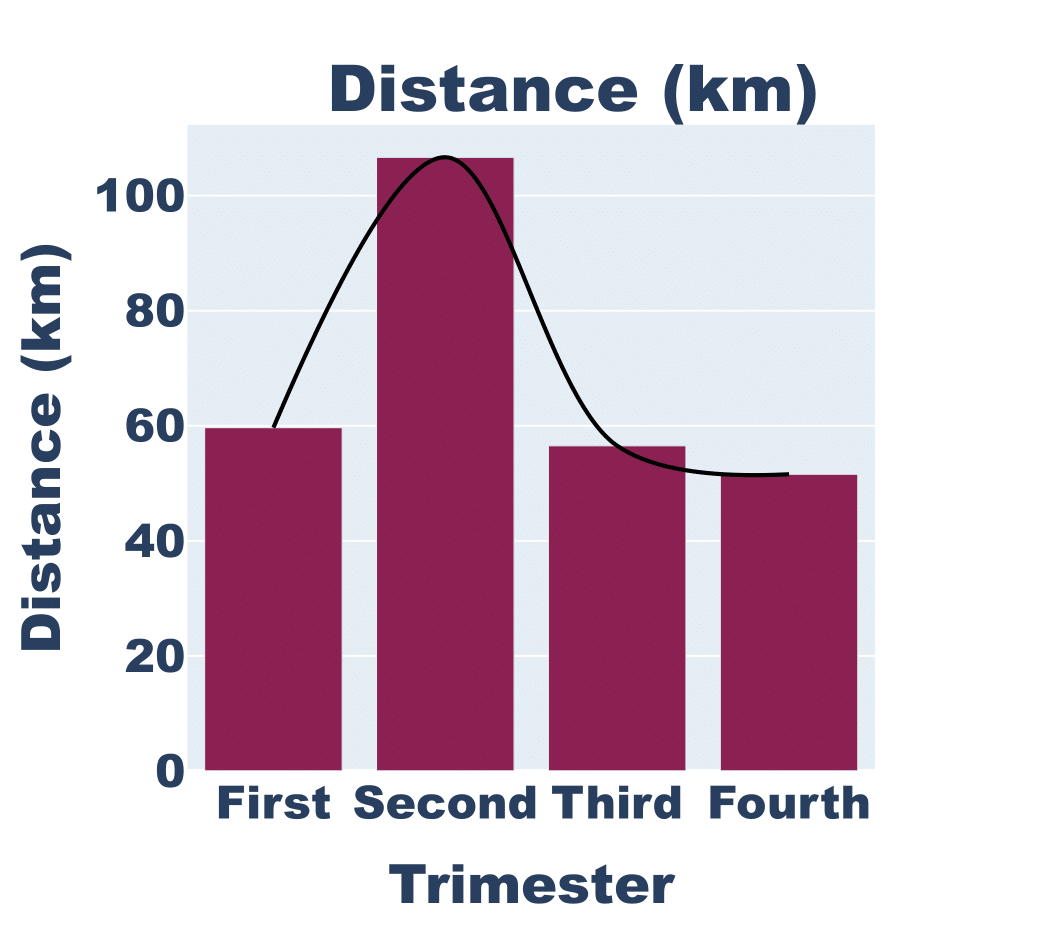}\\
    \includegraphics[width=0.3\textwidth]{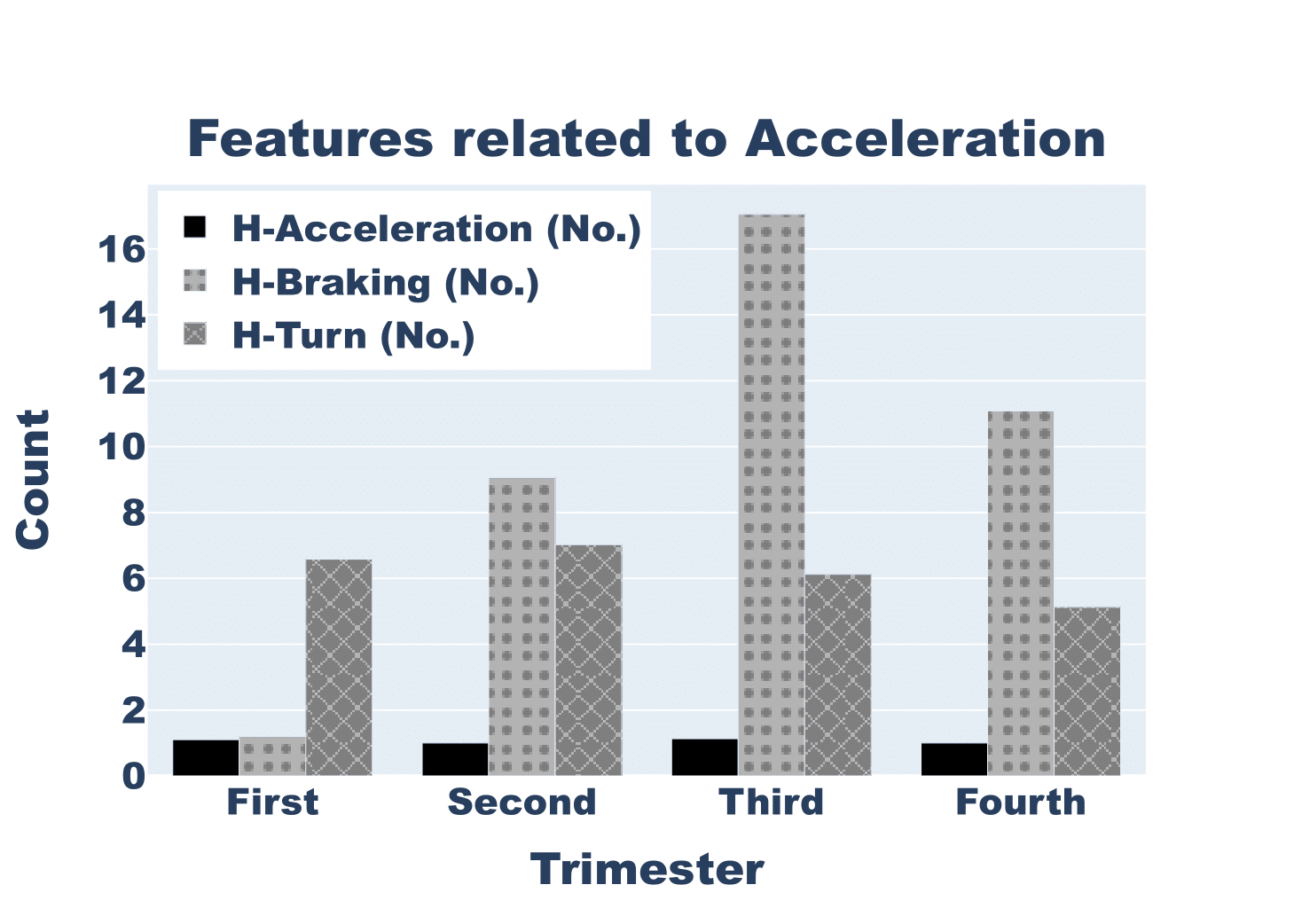}
    \vline
    \includegraphics[width=0.3\textwidth]{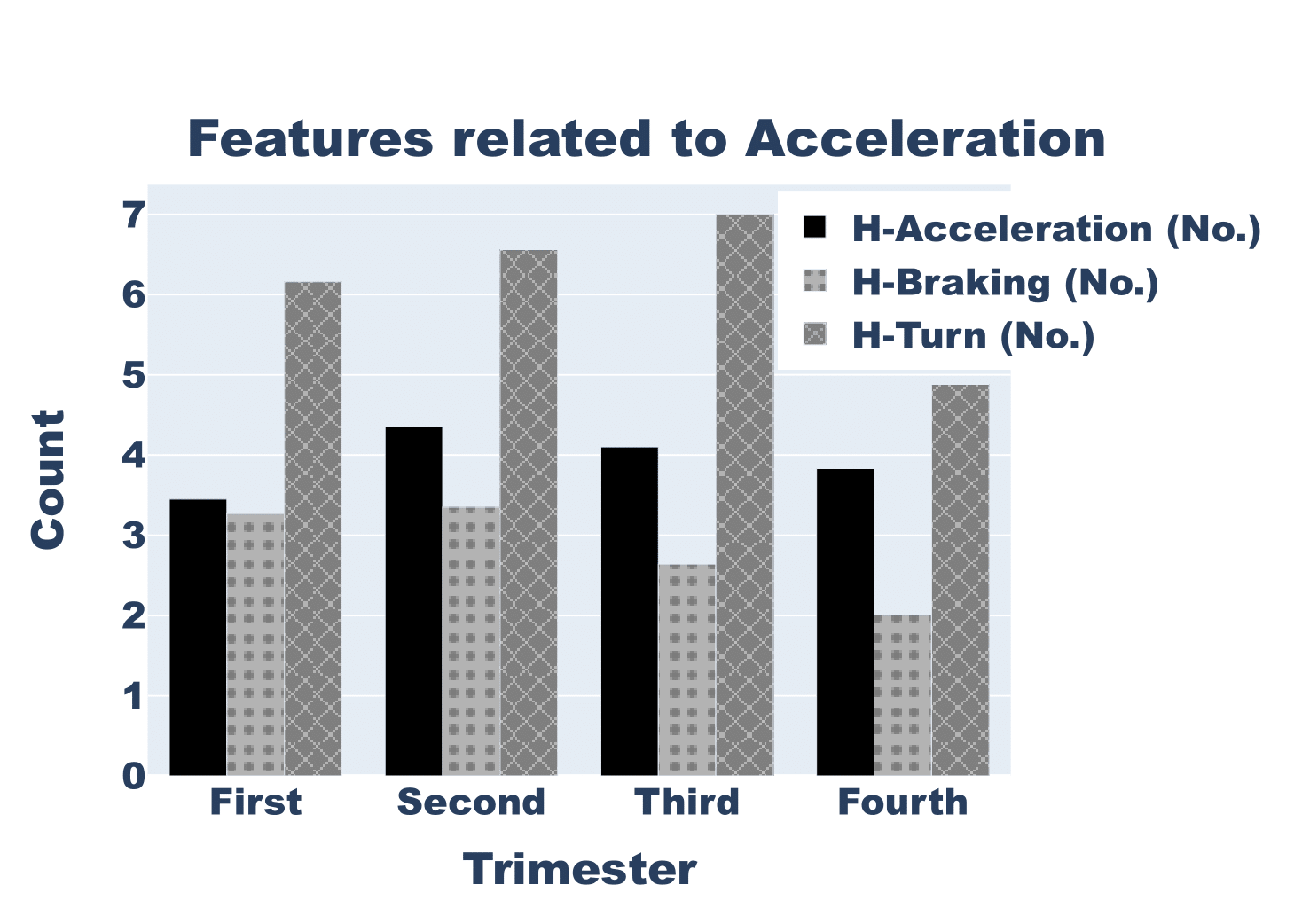}\\
    \includegraphics[width=0.3\textwidth]{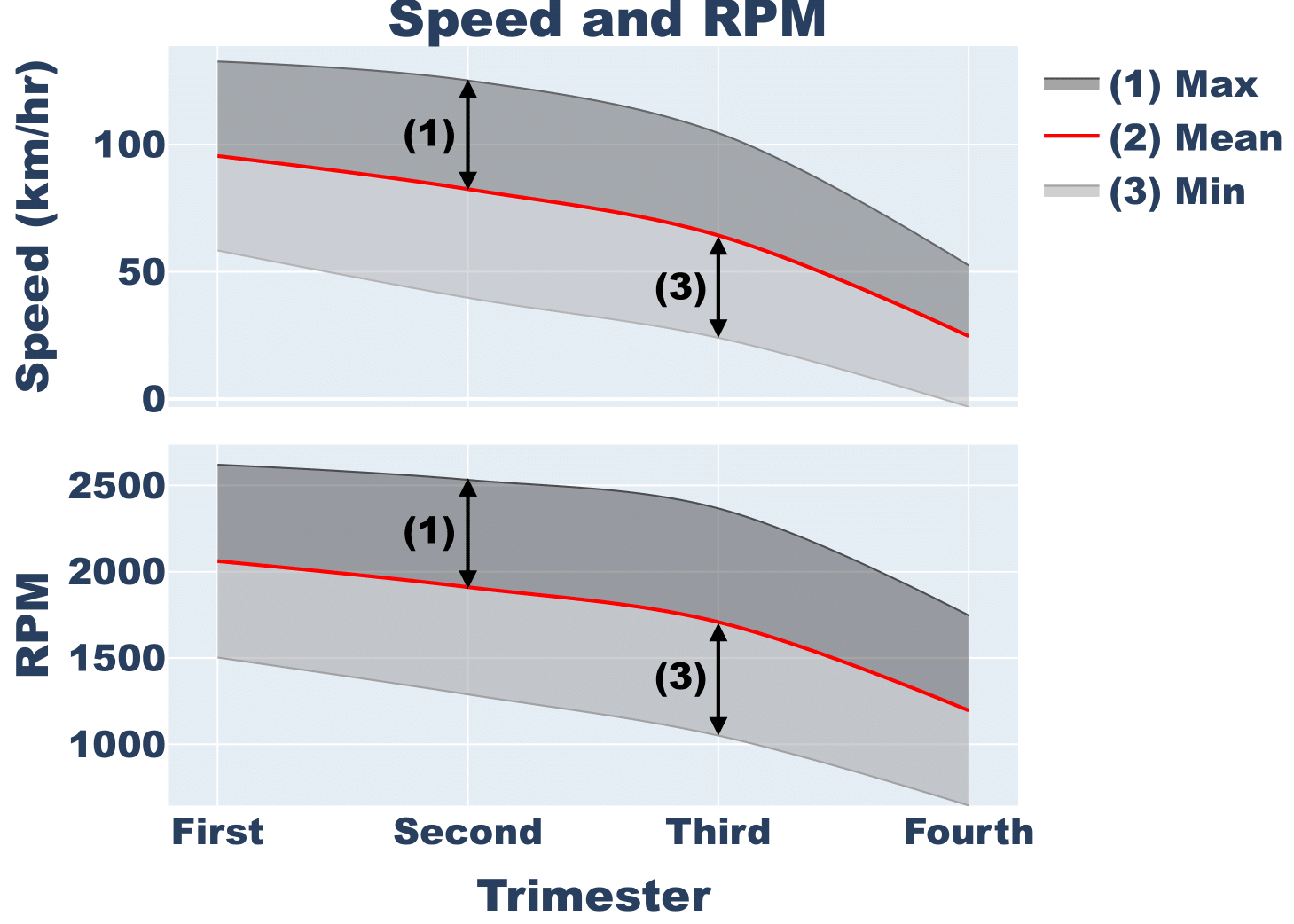}~(a)
    \vline
    ~(b){\includegraphics[width=0.3\textwidth]{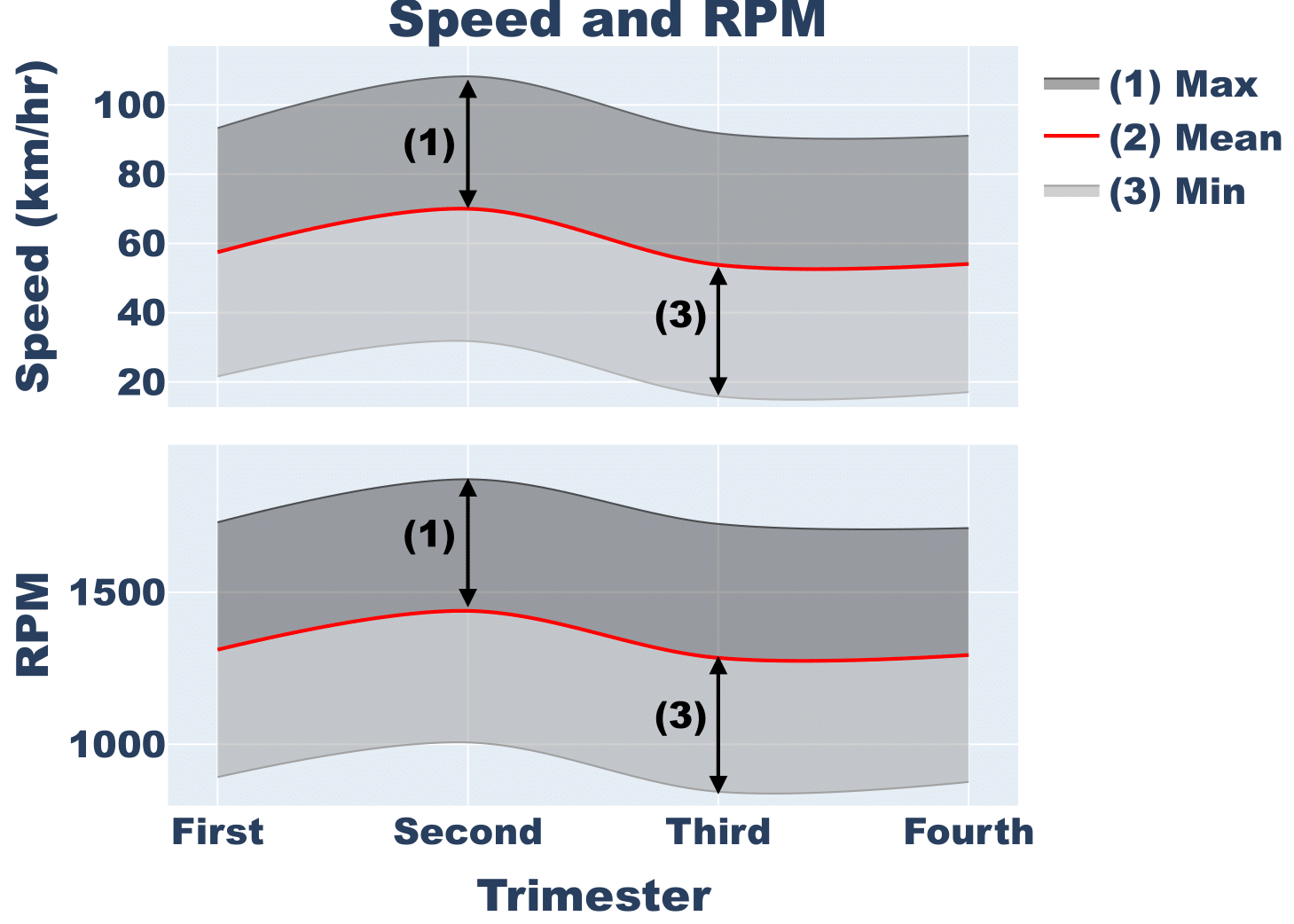}
    \caption{Quarterly Evaluation Reports ((a)-non-MCI; (b)-MCI)}
    \label{fig4}
    }
\end{figure*}

\section{Results}
\subsection{Statististical reports}
We analyzed the time of day for a subset of the participants. We divided the time into four categories: 5:00 to 11:59 as the morning; 12:00 to 16:59 as the afternoon; 17:00 to 20:59 as the evening; and 21:00 to 4:59 as the night. Fig.~\ref{fig3} represents the number of trips in each category using a pie chart. According to the results, half of the trips occur between 12:00 and 16:59, meaning participants prefer to drive more in the afternoon and less at night. Considering 34.9 percent for morning trips, older drivers prefer driving during daylight. Fig.~\ref{fig4} depicts the three-month aggregated data for participants with and without MCI, which presents changes in their driving patterns during quarterly visits.

\subsection{Random Forests Model}
Table~\ref{table2} reports the performance metrics and confusion matrix for the developed model in six groups. The first model represents the contribution of age as the only variable for evaluation. The obtained accuracy needs to show better performance for this model. The second model considered only the number of trips for three categories: the total number of trips, the number of trips at night, and the number of trips for peak hours. This model offers a better performance in comparison to the first model. Model 4 is related to only drivers and their demographic characteristics. This model evaluates age, gender, ethnicity, education, race, body mass index (BMI), and employment status. The performance result is approximately similar to the first model. Models 4, 5, and 6 recruit driving variables with three different preferences involving driver variables with orders of no driver variables, only age, and all driver variables. All these three models show a similar performance as a result.

\begin{table}[]
\fontsize{18pt}{18pt}\selectfont
\caption{Evaluation of Random Forest Model Performance Metrics}
\label{table2}
\resizebox{\columnwidth}{!}{%
\begin{tabular}{cccccccccc}
Model & Input & Accuracy & AUC & Precision & Recall & {\begin{tabular}[c]{@{}c@{}}F1\\ Score\end{tabular}} & \multicolumn{3}{c}{\begin{tabular}[c]{@{}c@{}}Confusion\\ Matrix\end{tabular}} \\
\hline
      &       &          &     &           &        &          & Predicted & \multicolumn{2}{c}{Observed}  \\
      &       &          &     &           &        &          &           & 0 & 1 \\
\hline      
1     & Only age & 0.79 & 0.87 & 1.00 & 0.70 & \multicolumn{1}{c|}{0.82} & \multicolumn{1}{c|}{0} & 1242 & 533 \\
      &          &      &      & 0.60 & 1.00 & \multicolumn{1}{c|}{0.75} & \multicolumn{1}{c|}{1} & 0    & 797 \\
2     & \begin{tabular}[c]{@{}c@{}}Number of trips\\ (total, peak, night)\end{tabular} & 0.86 & 0.91 & 0.83 & 1.00 & \multicolumn{1}{c|}{0.91} & \multicolumn{1}{c|}{0} & 1775 & 0 \\
      &          &      &      & 1.00 & 0.54 & \multicolumn{1}{c|}{0.70} & \multicolumn{1}{c|}{1} & 366  & 431 \\
3     & Driver Variables & 0.82 & 0.87 & 0.79 & 1.00 & \multicolumn{1}{c|}{0.88} & \multicolumn{1}{c|}{0} & 1774 & 1 \\
      &                  &      &      & 1.00 & 0.41 & \multicolumn{1}{c|}{0.58} & \multicolumn{1}{c|}{1} & 468  & 329 \\
4     & Driving Variables & 0.78 & 0.93 & 0.75 & 1.00 & \multicolumn{1}{c|}{0.86} & \multicolumn{1}{c|}{0} & 1775 & 0 \\
      &                   &      &      & 1.00 & 0.28 & \multicolumn{1}{c|}{0.43} & \multicolumn{1}{c|}{1} & 576  & 221 \\
5     & Age with driving variables & 0.86 & 0.94 & 0.83 & 1.00 & \multicolumn{1}{c|}{0.91} & \multicolumn{1}{c|}{0} & 1774 & 1 \\
      &                            &      &      & 1.00 & 0.54 & \multicolumn{1}{c|}{0.70} & \multicolumn{1}{c|}{1} & 366  & 431 \\
6     & All the variables & 0.86 & 0.95 & 0.83 & 1.00 & \multicolumn{1}{c|}{0.91} & \multicolumn{1}{c|}{0} & 1775 & 0 \\
      &                   &      &      & 1.00 & 0.54 & \multicolumn{1}{c|}{0.70} & \multicolumn{1}{c|}{1} & 366  & 431 \\
\end{tabular}%
}
\end{table}

\begin{figure}[htbp]
\centerline{\includegraphics[width=\columnwidth]{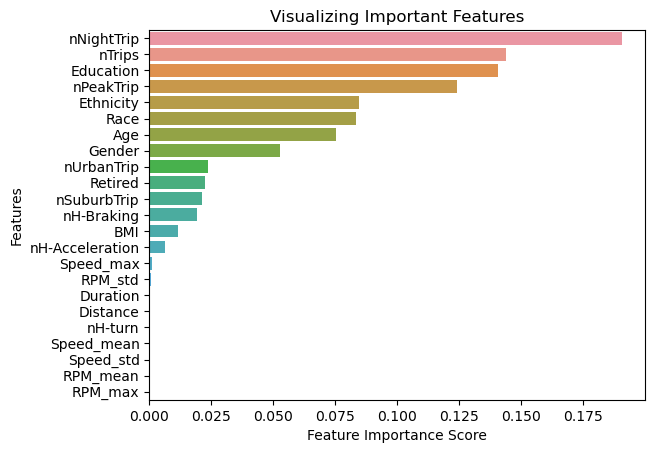}}
\vspace{-10pt}
\caption{Importance of Features}
\label{fig5}
\end{figure}

\section{Conclusion}

This research addresses the pressing problem of identifying and assessing mild cognitive impairment (MCI) among older drivers using in-vehicle sensing technology and machine learning methods. This study aims to develop a system capable of recognizing early signs of cognitive decline and providing appropriate interventions to ensure the safety and well-being of older drivers. Integrating sensors such as cameras, accelerometers, and the telematics unit into vehicles enables the collection of various data points related to driving behavior, vehicle dynamics, and driver interactions. These data points serve as valuable indicators, helping to reveal subtle alterations associated with MCI. The potential benefits of these systems are undeniable. Early identification of MCI can enable timely intervention and support for older drivers, improving safety while prolonging independence and mobility. By harnessing in-vehicle sensing technology and machine learning techniques, this study opens the way to developing intelligent systems capable of responding in real-time to diverse driving scenarios and detecting and assessing MCI.

This system can classify and predict MCI in real time by training machine learning models on diverse datasets. These models may provide timely interventions, such as alerts or adaptive driving assistance, to support older drivers with MCI and mitigate potential road risks. Machine learning techniques offer several advantages in this respect, allowing for the creation of models that can recognize anomalous patterns or conditions indicative of MCI, adapt, and improve over time as more data is collected and analyzed. In addition, quickly processing large amounts of data enables real-time monitoring and decision-making that allows prompt responses to changing cognitive states. Depending on the training data's quality, there can be various challenges and considerations related to the accuracy and reliability of machine learning models. Collecting diverse and comprehensive data sets accurately representing MCI-related driving behaviors and patterns is protracted. Ensuring the privacy and security of the data collection process is also essential. Measures taken must anonymize and protect sensitive personal data while permitting practical analysis.

Overall, this study aims to assist in designing advanced assisting systems for supporting older drivers while driving. The findings provide insight into novel solutions that improve road safety and quality of life for older drivers with mild cognitive impairments.

\section*{Acknowledgment}

We want to express our gratitude to the U.S. National Institute of Health (NIH) (R01 AG068472 02S1) and the National Science Foundation (NSF) (OAC 1948066) for their excellent assistance and financing.

\bibliographystyle{ieeetr}

\bibliography{citation}

\begin{thebibliography}{10}
\providecommand{\url}[1]{#1}
\csname url@samestyle\endcsname
\providecommand{\newblock}{\relax}
\providecommand{\bibinfo}[2]{#2}
\providecommand{\BIBentrySTDinterwordspacing}{\spaceskip=0pt\relax}
\providecommand{\BIBentryALTinterwordstretchfactor}{4}
\providecommand{\BIBentryALTinterwordspacing}{\spaceskip=\fontdimen2\font plus
\BIBentryALTinterwordstretchfactor\fontdimen3\font minus \fontdimen4\font\relax}
\providecommand{\BIBforeignlanguage}[2]{{%
\expandafter\ifx\csname l@#1\endcsname\relax
\typeout{** WARNING: IEEEtran.bst: No hyphenation pattern has been}%
\typeout{** loaded for the language `#1'. Using the pattern for}%
\typeout{** the default language instead.}%
\else
\language=\csname l@#1\endcsname
\fi
#2}}
\providecommand{\BIBdecl}{\relax}
\BIBdecl

\bibitem{he2016aging}
W.~He, D.~Goodkind, P.~R. Kowal \emph{et~al.}, ``An aging world: 2015,'' 2016.

\bibitem{united2021world}
\BIBentryALTinterwordspacing
U.~Nations, \emph{World Population Ageing 2020 Highlights: Living Arrangements of Older Persons}, ser. Economic \& social affairs.\hskip 1em plus 0.5em minus 0.4em\relax UN, 2021. [Online]. Available: \url{https://books.google.com/books?id=ApggzgEACAAJ}
\BIBentrySTDinterwordspacing

\bibitem{seifallahi2022detection}
M.~Seifallahi, J.~E. Galvin, and B.~Ghoraani, ``Detection of mild cognitive impairment from quantitative analysis of timed up and go (tug),'' in \emph{2022 IEEE International Conference on Data Mining Workshops (ICDMW)}.\hskip 1em plus 0.5em minus 0.4em\relax IEEE, 2022, pp. 248--253.

\bibitem{us2016vha}
U.~D. of~Veterans Affairs (VA) Veterans Health Administration (VHA) Dementia Steering Committee~(DSC), ``Vha dementia steering committee recommendations for dementia care in the vha healthcare system 2016,'' 2016.

\bibitem{altaher2020using}
A.~Altaher, Z.~Salekshahrezaee, A.~Abdollah~Zadeh, H.~Rafieipour, and A.~Altaher, ``Using multi-inception cnn for face emotion recognition,'' \emph{Journal of Bioengineering Research}, vol.~3, no.~1, pp. 1--12, 2020.

\bibitem{brown2004driving}
L.~B. Brown and B.~R. Ott, ``Driving and dementia: a review of the literature,'' \emph{Journal of geriatric Psychiatry and Neurology}, vol.~17, no.~4, pp. 232--240, 2004.

\bibitem{gold2012examination}
D.~A. Gold, ``An examination of instrumental activities of daily living assessment in older adults and mild cognitive impairment,'' \emph{Journal of clinical and experimental neuropsychology}, vol.~34, no.~1, pp. 11--34, 2012.

\bibitem{rajan2013disability}
K.~B. Rajan, L.~E. Hebert, P.~A. Scherr, C.~F. Mendes~de Leon, and D.~A. Evans, ``Disability in basic and instrumental activities of daily living is associated with faster rate of decline in cognitive function of older adults,'' \emph{Journals of Gerontology Series A: Biomedical Sciences and Medical Sciences}, vol.~68, no.~5, pp. 624--630, 2013.

\bibitem{teasdale2016older}
N.~Teasdale, M.~Simoneau, L.~Hudon, M.~Germain~Robitaille, T.~Moszkowicz, D.~Laurendeau, L.~Bherer, S.~Duchesne, and C.~Hudon, ``Older adults with mild cognitive impairments show less driving errors after a multiple sessions simulator training program but do not exhibit long term retention,'' \emph{Frontiers in human neuroscience}, vol.~10, p. 653, 2016.

\bibitem{croston2009driving}
J.~Croston, T.~M. Meuser, M.~Berg-Weger, E.~A. Grant, and D.~B. Carr, ``Driving retirement in older adults with dementia,'' \emph{Topics in Geriatric Rehabilitation}, vol.~25, no.~2, p. 154, 2009.

\bibitem{kaye2011intelligent}
J.~A. Kaye, S.~A. Maxwell, N.~Mattek, T.~L. Hayes, H.~Dodge, M.~Pavel, H.~B. Jimison, K.~Wild, L.~Boise, and T.~A. Zitzelberger, ``Intelligent systems for assessing aging changes: home-based, unobtrusive, and continuous assessment of aging,'' \emph{Journals of Gerontology Series B: Psychological Sciences and Social Sciences}, vol.~66, no. suppl\_1, pp. i180--i190, 2011.

\bibitem{lyons2015pervasive}
B.~E. Lyons, D.~Austin, A.~Seelye, J.~Petersen, J.~Yeargers, T.~Riley, N.~Sharma, N.~Mattek, K.~Wild, H.~Dodge \emph{et~al.}, ``Pervasive computing technologies to continuously assess alzheimer’s disease progression and intervention efficacy,'' \emph{Frontiers in aging neuroscience}, vol.~7, p. 102, 2015.

\bibitem{teipel2018use}
S.~Teipel, A.~K{\"o}nig, J.~Hoey, J.~Kaye, F.~Kr{\"u}ger, J.~M. Robillard, T.~Kirste, and C.~Babiloni, ``Use of nonintrusive sensor-based information and communication technology for real-world evidence for clinical trials in dementia,'' \emph{Alzheimer's \& Dementia}, vol.~14, no.~9, pp. 1216--1231, 2018.

\bibitem{shuqair2022incremental}
M.~Shuqair, J.~Jimenez-Shahed, and B.~Ghoraani, ``Incremental learning in time-series data using reinforcement learning,'' in \emph{2022 IEEE International Conference on Data Mining Workshops (ICDMW)}.\hskip 1em plus 0.5em minus 0.4em\relax IEEE, 2022, pp. 868--875.

\bibitem{seelye2017passive}
A.~Seelye, N.~Mattek, N.~Sharma, I.~Witter, A.~Brenner, K.~Wild, H.~Dodge, J.~Kaye \emph{et~al.}, ``Passive assessment of routine driving with unobtrusive sensors: A new approach for identifying and monitoring functional level in normal aging and mild cognitive impairment,'' \emph{Journal of Alzheimer's disease}, vol.~59, no.~4, pp. 1427--1437, 2017.

\bibitem{jan2023methods}
M.~T. Jan, S.~Moshfeghi, J.~W. Conniff, J.~Jang, K.~Yang, J.~Zhai, M.~Rosselli, D.~Newman, R.~Tappen, and B.~Furht, ``Methods and tools for monitoring driver's behavior,'' \emph{arXiv preprint arXiv:2301.12269}, 2023.

\bibitem{jan2022non}
M.~T. Jan, A.~Hashemi, J.~Jang, K.~Yang, J.~Zhai, D.~Newman, R.~Tappen, and B.~Furht, ``Non-intrusive drowsiness detection techniques and their application in detecting early dementia in older drivers,'' in \emph{Proceedings of the Future Technologies Conference (FTC) 2022, Volume 2}.\hskip 1em plus 0.5em minus 0.4em\relax Springer, 2022, pp. 776--796.

\bibitem{adler2005older}
G.~Adler, S.~Rottunda, and M.~Dysken, ``The older driver with dementia: an updated literature review,'' \emph{Journal of safety research}, vol.~36, no.~4, pp. 399--407, 2005.

\bibitem{li2017longitudinal}
G.~Li, D.~W. Eby, R.~Santos, T.~J. Mielenz, L.~J. Molnar, D.~Strogatz, M.~E. Betz, C.~DiGuiseppi, L.~H. Ryan, V.~Jones \emph{et~al.}, ``Longitudinal research on aging drivers (longroad): study design and methods,'' \emph{Injury Epidemiology}, vol.~4, pp. 1--16, 2017.

\bibitem{khademi2022comprehensive}
S.~Khademi, M.~Neghabi, M.~Farahi, M.~Shirzadi, and H.~R. Marateb, ``A comprehensive review of the movement imaginary brain-computer interface methods: Challenges and future directions,'' \emph{Artificial Intelligence-Based Brain-Computer Interface}, pp. 23--74, 2022.

\bibitem{neghabi2022novel}
M.~Neghabi, H.~R. Marateb, and A.~Mahnam, ``Novel frequency-based approach for detection of steady-state visual evoked potentials for realization of practical brain computer interfaces,'' \emph{Brain-Computer Interfaces}, vol.~9, no.~3, pp. 155--168, 2022.

\bibitem{alkanjr2023iobt}
B.~Alkanjr and T.~Alshammari, ``Iobt intrusion detection system using machine learning,'' in \emph{2023 IEEE 13th Annual Computing and Communication Workshop and Conference (CCWC)}.\hskip 1em plus 0.5em minus 0.4em\relax IEEE, 2023, pp. 0886--0892.

\bibitem{lyu2022using}
N.~Lyu, Y.~Wang, C.~Wu, L.~Peng, and A.~F. Thomas, ``Using naturalistic driving data to identify driving style based on longitudinal driving operation conditions,'' \emph{Journal of Intelligent and Connected Vehicles}, vol.~5, no.~1, pp. 17--35, 2022.

\bibitem{bayat2021gps}
S.~Bayat, G.~M. Babulal, S.~E. Schindler, A.~M. Fagan, J.~C. Morris, A.~Mihailidis, and C.~M. Roe, ``Gps driving: a digital biomarker for preclinical alzheimer disease,'' \emph{Alzheimer's Research \& Therapy}, vol.~13, no.~1, pp. 1--9, 2021.

\bibitem{eby2012driving}
D.~W. Eby, N.~M. Silverstein, L.~J. Molnar, D.~LeBlanc, and G.~Adler, ``Driving behaviors in early stage dementia: A study using in-vehicle technology,'' \emph{Accident Analysis \& Prevention}, vol.~49, pp. 330--337, 2012.

\bibitem{konjalwar2023demonstrating}
S.~Konjalwar, B.~Ceyhan, O.~Rivera, P.~Nategh, M.~Neghabi, M.~Pavlovic, S.~Allani, and M.~Ranji, ``Demonstrating drug treatment efficacies by monitoring superoxide dynamics in human lung cancer cells with time-lapse fluorescence microscopy,'' 2023.

\bibitem{farsi1999overview}
M.~Farsi, K.~Ratcliff, and M.~Barbosa, ``An overview of controller area network,'' \emph{Computing \& Control Engineering Journal}, vol.~10, no.~3, pp. 113--120, 1999.

\bibitem{johansson2005vehicle}
K.~H. Johansson, M.~T{\"o}rngren, and L.~Nielsen, ``Vehicle applications of controller area network,'' in \emph{Handbook of networked and embedded control systems}.\hskip 1em plus 0.5em minus 0.4em\relax Springer, 2005, pp. 741--765.

\bibitem{leonhardt2018unobtrusive}
S.~Leonhardt, L.~Leicht, and D.~Teichmann, ``Unobtrusive vital sign monitoring in automotive environments—a review,'' \emph{Sensors}, vol.~18, no.~9, p. 3080, 2018.

\bibitem{shrestha2017factors}
P.~P. Shrestha and K.~J. Shrestha, ``Factors associated with crash severities in built-up areas along rural highways of nevada: A case study of 11 towns,'' \emph{Journal of traffic and transportation engineering (English edition)}, vol.~4, no.~1, pp. 96--102, 2017.

\end{thebibliography}

\end{document}